\documentclass[preprint]{aastex}
\def\stacksymbols #1#2#3#4{\def\theguybelow{#2}
        \def\verticalposition{\lower#3pt}
        \def\spacingwithinsymbol{\baselineskip0pt\lineskip#4pt}
        \mathrel{\mathpalette\intermediary#1}}
\def\intermediary #1#2{\verticalposition\vbox{\spacingwithinsymbol
        \everycr={}\tabskip0pt
        \halign{$\mathsurround0pt#1\hfil##\hfil$\crcr#2\crcr
                \theguybelow\crcr}}}
\def\lta{\stacksymbols{<}{\sim}{2.5}{.2}}
\def\gta{\stacksymbols{>}{\sim}{3}{.5}}

\begin{document}

\title{THERMAL EVOLUTION OF SUPERNOVA IRON IN ELLIPTICAL GALAXIES}

\author{ Fabrizio Brighenti$^{1,2}$ \& William G. Mathews$^1$}

\affil{$^1$University of California Observatories/Lick Observatory,
Department of Astronomy and Astrophysics,
University of California, Santa Cruz, CA 95064\\
mathews@ucolick.org}

\affil{$^2$Dipartimento di Astronomia,
Universit\`a di Bologna,
via Ranzani 1,
Bologna 40127, Italy\\
brighenti@bo.astro.it}

\begin{abstract}
Interpretations of the spatial distribution, 
abundance ratios and global masses of metals 
in the hot gas of galaxy clusters in terms of 
supernova enrichment have been problematical. 
For example, the abundance of iron and other elements 
occasionally declines toward the center 
just where the stellar and supernova densities are highest.
Also, the mass of gas-phase iron per unit stellar mass or light
is lower in elliptical galaxies and groups than in 
rich galaxy clusters. 
We discuss hypothetical scenarios in which these abundance 
anomalies can result from the preferential buoyant separation 
of metals. 
However, in this and 
all previous attempts to explain these metallicity observations 
it has been assumed that all metals created by supernovae
are present either in visible stars or the hot gas. 
We discuss here the possibility that some of the iron 
expelled into the hot gas by Type Ia supernovae  
may have radiatively cooled, avoiding detection by 
X-ray and optical observers. 
Hydrodynamic models of Type Ia explosions in the hot gas inside 
elliptical galaxies create a gas of nearly pure iron that is  
several times hotter than the local interstellar gas. 
We describe the subsequent thermal evolution of the iron-rich 
gas as it radiates and thermally mixes with the surrounding gas.
There is a critical time by which the iron ions must mix into 
the ambient gas to avoid rapid radiative cooling.
We find that successful mixing is possible if 
the iron ions diffuse with large mean free paths, 
as in an unmagnetized plasma. 
However, the Larmor radii of the iron ions are exceptionally small 
in microgauss fields, so the field geometry must be highly 
tangled or radial 
to allow the iron to mix by diffusion faster than 
it cools by radiative losses. 
The possibility that some of the supernova iron cools cannot be 
easily discounted.
\end{abstract}


\section{Introduction}

We explore the possibility that some iron produced by 
Type Ia supernovae in galaxy groups and clusters 
may have radiatively cooled and avoided detection 
by X-ray and optical observers. 
Because of the large radiative emissivity 
of extremely iron-rich plasmas produced by 
Type Ia supernovae (SNIae), 
it is possible that this gas cools rapidly 
to low temperatures where it may be difficult 
to observe. 
Even if all SNIa iron 
radiatively cooled in an elliptical galaxy,
the total collective X-ray luminosity 
would be small because the thermal energy 
content of the iron-rich cores in 
Type Ia supernova remnants is very much 
less than that of the radiating hot gas distributed throughout 
the galaxy or cluster. 

Another motivation for this inquiry 
are the central abundance minima often observed in the hot gas
inside central massive ellipticals in galaxy groups and clusters
(Dupke \& White, 2000;
Johnstone, et al 2002;
Sanders \& Fabian 2002;
Schmidt, Fabian \& Sanders 2002;
Blanton, Sarazin \& McNamara 2003;
Buote et al. 2003).
When present, these minima typically lie within
the central $\lta 30$ kpc, 
just where abundances in the hot gas
are likely to be 
dominated by SNIae in the central elliptical galaxy.
(The stellar density is more centrally peaked
than the gas density.)

Another possible motivation is the observation 
of unexpectedly low iron abundance 
in the hot gas within some elliptical galaxies
of moderate or low $L_B$.
In a recent study of three X-ray faint elliptical
galaxies
O'Sullivan \& Ponman (2004) found very low
iron abundances in the interstellar gas, 
$z_{Fe} \lta 0.1z_{Fe\odot}$. 
Using {\it Chandra} data 
Humphrey, Buote \& Canizares (2004) 
estimated $z_{Fe} \sim z_{Fe\odot}$ in the low
$L_x/L_B$ early-type galaxies NGC 1332 and NGC 720,
perhaps indicating that $\sim$ solar gas-phase iron abundances
are more typical.
However, the supersolar iron abundances
predicted by our gasdynamical models for isolated 
elliptical galaxies typically exceed $z_{Fe\odot}$
(e.g. Mathews \& Brighenti 2003).

Additional motivations for considering 
the cooling of iron-rich supernova remnants 
are (1) the difficulty of understanding how 
Type II and Type Ia supernovae enrich the hot 
gas in rich clusters of galaxies
(Gibson \& Matteucci 1997; Moretti et al. 2003;
Baumgartner et al. 2003; 
Loewenstein 2004; Portinari et al. 2004;
Tornatore et al. 2004, but see Tamura et al. 2004) 
and (2) the discord 
in metal abundances between rich galaxy clusters 
and galaxy groups or elliptical galaxies
(e.g. Renzini 1997, 2000; Brighenti \& Mathews 1999). 
Perhaps there are significant errors in supernova yields or 
in metallicity observations. 

But in these 
recent attempts to understand the metal enrichment history of 
galaxy clusters,
it is assumed that all of the iron produced by 
supernovae is still currently visible either in 
stars or in the hot gas. 
This is the assumption that we investigate here.
It is necessary to establish that the iron visible
today faithfully represents the total iron produced over time,
particularly since the total historical 
number of supernovae in old
stellar populations is uncertain.
Alternatively, the lower iron content 
in groups and individual elliptical galaxies
relative to massive clusters 
may be explained if supernova products are 
physically removed 
from the observed hot gas by winds or buoyancy.
This scenario could simultaneously explain 
the relatively low iron abundance in gas associated 
with individual non-cluster elliptical galaxies 
and the larger amount of gas-phase iron in clusters.

Morris \& Fabian (2003) show that the enhanced radiative cooling 
of iron-rich inhomogeneities may occur without producing appreciable 
evidence of cooling in X-ray spectra. 
Unlike the iron-rich cooling discussed by Morris \& Fabian
-- in which gas with different abundances was assumed to be 
at the same initial temperature -- the iron in SNIa 
remnants is created 
at temperatures that significantly exceed that of the ambient hot gas, 
so the removal of SNIa iron by selective radiative cooling is not 
immediately obvious. 
In view of this we also estimate the inhomogeneity in 
SNIa iron that would be
expected due to the sparse distribution of SNIa
events on galactic scales.

In the following we describe possible thermal histories for 
the iron produced by SNIa explosions in elliptical galaxies.
We begin with estimates of the post-supernova 
temperature of the iron-rich gas using gasdynamical models.
This is followed by a discussion of possible scenarios for the 
rate that the iron-rich gas cools by thermal mixing and 
radiation losses. 
Our results suggest that much of the SNIa iron 
in cluster-centered elliptical galaxies may 
mix successfully with the ambient hot gas, but only 
if magnetic fields have little influence on the 
diffusion of iron ions.

\section{Evolution of SNIa Remnants in Hot Gas}

As we discuss below, the long term survival of iron 
introduced by Type Ia supernovae into the hot interstellar 
gas depends critically on the additional 
thermal energy imparted to the 
gas during the supernova explosion. 
Most calculations of the evolution of supernova remnants
describe the early, luminous stages relevant
to remnants observed in the Milky Way. 
Instead, we are interested in estimating the
temperature of the iron-rich remnant gas after $10^4 - 10^5$ years 
when it has acquired essentially the same pressure as the local 
interstellar gas in a typical elliptical galaxy. 
The energy released by the $\beta$ decay of cobalt, 
which sustains the SNIa light curve and decreases exponentially 
$\sim e^{-t/t_{\beta}}$ in $t_{\beta} \approx 100$ days, 
can be neglected in the total energy budget and 
on the very long time scales of interest here.

The one dimensional gas dynamical calculations of 
Mathews (1990) and Dorfi \& V\"olk (1996) 
describe various aspects of Type Ia explosions in 
the rarefied gas in elliptical galaxies. 
The expanding stellar remnant cools 
by adiabatic expansion to exceedingly low 
temperatures and is then heated 
by a reverse shock as it is decelerated by 
the interstellar gas. 
>From dimensional considerations, the characteristic temperature of 
the stellar iron is 
$T \sim (E_{sn}/M_{sn})m_p/k \sim 4 \times 10^9$ K 
for $E_{sn} = 10^{51}$ ergs and $M_{sn} = 1.4$ $M_{\odot}$. 
But this greatly overestimates the final iron temperature because 
(1) only a rather small amount 
of gas experiences shock velocities as large as  
$\sim (E_{sn}/M_{sn})^{1/2}$, 
and (2) only a comparatively 
small fraction of the supernova energy goes into heating 
the stellar gas in the stellar remnant -- most of the 
supernova energy is delivered 
to the surrounding interstellar medium 
by an expanding shock wave.

To estimate the final temperature 
of the iron-rich gas 
we have calculated several simple one dimensional 
SNIa blast waves 
in the hot interstellar gas in elliptical galaxies.
It is generally accepted that 
small particle Larmor radii in microgauss 
magnetic fields allow these strong shocks 
to be treated with normal continuum 
gas dynamics and we assume this to be true. 
We also assume spherical symmetry which is unlikely to 
hold in detail.
Non-spherical irregularities resulting from 
subsonic deflagration are expected inside the young 
stellar remnant and the deceleration of the 
expanding stellar gas results in Rayleigh-Taylor 
instabilities producing further spatial deformations. 
However, these details, which are ignored here, are unlikely 
to significantly influence our
estimates of the mean gas temperature 
during the final stages of the remnant evolution 
when hydrostatic equilibrium is established throughout the 
iron-rich gas.

For an initial remnant configuration 
we employ the simple post-deflagration stellar remnant 
described for the W7 SNIa model of mass $M_{sn} =1.4$ 
$M_{\odot}$ proposed by 
Dwarkadas \& Chevalier (1998), which is based on the calculations by
H\"oflich \& Khokhlov (1996). 
The density distribution at time $t_i$ is described by 
\begin{displaymath}
\rho = \rho_o e^{-(r/r_s)} ~~~{\rm gm~cm}^{-3}
\end{displaymath}
with $r_s = v_s t_i$.
The initial velocity field is linear $v = v_s r/r_s$.
For any assumed initial 
time $t_i$ integrations of $\rho(r)$ 
to a total mass $M_{sn} = 1.4$ $M_{\odot}$ and the kinetic energy density 
$\rho v_s^2/2$ to 
$E_{sn} = 10^{51}$ ergs determine $\rho_o$ and 
the initial stellar radius $r_s$ 
of the contact discontinuity between stellar and ambient gas.
The initial time $t_i = 10^7$ sec for our calculations is chosen 
so that $r_s$ is much less than the radius of ambient gas 
that contains the stellar mass ejected 
$r_m = (3 M_{sn} / 4 \pi \rho)^{1/3} = 2.27/n_e^{1/3}$ pc 
where $\rho = 1.17 n_e m_p$ and $n_e$ are the ambient 
interstellar mass and electron densities. 

To estimate the temperature of the iron plasma when the 
final SNIa remnant has reached pressure equilibrium, 
we calculated the evolution of several 
remnants for $\sim 10^5$ years 
using a high-resolution 1D Eulerian code.
For simplicity 
the entire $1.4 M_{\odot}$ star is assumed to be essentially 
pure iron ($z_{Fe} = 10^5z_{Fe\odot}$);
the hot ambient gas is assumed to have solar abundance. 
The code solves an additional equation 
for the iron density to determine the iron abundance 
$z_{Fe}(r,t)$. 
In Figure 1 we illustrate the gas temperature profile $T(r)$
after time $t = 5 \times 10^4$ years produced in 
interstellar medium of density $n_e = 0.01$ cm$^{-3}$ and temperature 
$T = 10^7$ K. 
The vertical dashed line in Figure 1 shows the full extent of the 
stellar gas and the arrow points to the radius 
of the stellar core that includes
0.7 $M_{\odot}$ of pure iron. 
The outer layers of the expanded star contain Si 
and other elements.
At this time the stellar-interstellar contact discontinuity 
is at $\sim 19$ pc and the pressure is nearly uniform at 
all radii and equal to that in the initial ambient gas.
Beyond the iron rich remnant the gas temperature decreases, reflecting 
the weakening of the expanding blast wave.
The shock at $t = 5 \times 10^4$ yrs can be seen in Figure 1 
as a small vertical feature at 65 pc. 
The temperature of the iron-rich stellar gas 
generally increases toward the center, 
corresponding to converging reverse shocks 
that grow stronger with time.
An insignificant mass of gas near the center has a temperature 
comparable to $(E_{sn}/M_{sn})m_p/k$.

For the profile shown in Figure 1 we find that the mass weighted 
mean temperature in the central 0.7$M_{\odot}$ is 
$\langle T \rangle_{0.7} \approx 6.7 \times 10^7$ K. 
Table 1 lists the mean gas temperatures
$\langle T \rangle_{0.7}$ and
$\langle T \rangle_{0.35}$ within the central
$0.7$ and $0.35 M_{\odot}$ in the final
pressure-equilibrium remnants (at time $t = 5 \times 10^4$ yr)
computed for ambient gas
at several densities and temperatures typical of
hot gas at different radii inside elliptical galaxies.
All supernovae we consider have mass
$M_{sn} = 1.4 M_{\odot}$ and
energy $E_{sn} = 10^{51}$ ergs.
For interstellar gas at temperature $10^7$ K, the 
final equilibrium iron plasma is heated by a factor of 6 or 7, 
but this factor decreases with increasing initial temperature 
of the ambient gas.
In considering the subsequent thermal evolution of iron-rich 
gas in the following Section we consider  
$10^7 \lta \langle T \rangle \lta 3 \times 10^8$ K 
as representative of $\langle T \rangle$ values in Table 1.
In the discussion below we describe how this gas cools 
by mixing with ambient interstellar gas at $T = 10^7$ K 
and by optically thin radiative emission.

\section{Molecular Weight and Density}

We consider a collisionally ionized plasma 
of hydrogen, helium and iron. 
The temperature range of interest extends from $T \sim 10^5$ K, where
H and He are fully ionized, to $T \sim 3 \times 10^8$ K,
corresponding to completely ionized iron 
(Sutherland \& Dopita 1993). 
The molecular weight $\mu$ can be found from 
\begin{displaymath}
{1 \over \mu} = \sum_{j} {a_j \over A_j} n_{p,j}
= 2 a_H + {3 \over 4} a_{He} + {(1 + {\bar Z}) \over A}a_{Fe}.
\end{displaymath}
Here $n_{p,j}$ is the number of particles contributed by an atom of
element $j$. The fraction of mass in element $j$ is $a_j$ and
$A_j$ is its atomic number, 
where $A \equiv A_{Fe} = 56$ for iron.
Clearly $\sum a_j = 1$ and we adopt $a_{He}/a_H = 0.39$.
Combining these results, the molecular weight is
\begin{equation}
\mu = {{1.39 + z_{Fe}} \over {2.29 + z_{Fe}(1 + {\bar Z})/A}}
\end{equation}
where $z_{Fe} = a_{Fe}/a_H$ is the abundance of iron 
relative to hydrogen by weight 
and $a_{Fe}/(a_H + a_{He}) = z_{Fe}/1.39$. 
For a pure iron plasma, the molecular weight is simply 
$\mu_{Fe} = A/(1 + {\bar Z})$.

If $x_{Fe,i}(T)$ is the fraction of iron ions 
with charge $Z_i$, the mean charge of iron ions
\begin{equation}
{\bar Z}(T) = \sum_{i} x_{Fe,i} Z_i 
\end{equation}
varies slowly with temperature in collisional equilibrium.
In Figure 2 we plot ${\bar Z}(T)$ from 
Sutherland \& Dopita (1993) together with a simple analytic
fit
\begin{displaymath}
{\bar Z}(T) = [2.074(1 - \phi) + 3.387\phi]\log T 
\end{displaymath}
\begin{displaymath}
~~~~~~~~~~~~~+ [8.136(1 - \phi) - 13.12\phi]
\end{displaymath}
where
\begin{displaymath}
\phi = \{ 1 - \tanh [\log(T/10^{6.5}~\rm{K})/0.55] \}/2.
\end{displaymath}

For temperatures $10^5 \lta T \lta 10^8$ K, the 
energy radiated by line excitation and emission 
exceeds that in bremsstrahlung and other continua. 
In a plasma that cools by electron-ion collisions the
cooling rate depends critically on the 
number of free electrons per ion. 
At these temperatures 
only heavy ions such as iron have bound electrons. 
For our plasma of H, He and Fe, the number of 
free electrons per iron ion is 
\begin{equation}
{n_e \over n_{Fe}} = {{1.20 + z_{Fe} {\bar Z} /A} \over
{z_{Fe}/A}}.
\end{equation}
In a normal plasma with solar abundances 
($z_{Fe\odot} = 1.83 \times 10^{-3}$) 
the number of electrons per iron ion is 
$n_e/n_{Fe} \approx 1.20 A /z_{Fe\odot} = 3.67 \times 10^4$.
However, in a pure iron plasma ($z_{Fe} \rightarrow \infty$) 
this ratio 
$n_e/n_{Fe} = {\bar Z} \sim 21$ is very much reduced.
Since there are $\sim 10^3$ fewer electrons per iron ion available 
to excite iron transitions, the radiation efficiency 
of a pure iron plasma is less than might have been imagined. 
The critical iron abundance at which the number 
of electrons from H plus He equals that from iron is
\begin{equation}
z_{Fe,crit} \approx 1.20 A/{\bar Z} \approx 3.20 
= 1750 z_{Fe\odot}, 
\end{equation}
assuming ${\bar Z} \approx 21$.

The total density expressed in terms of number densities is 
\begin{equation}
\rho = m_p ( n_H + 4n_{He} + An_{Fe}) = n_H m_p [1.39 + z_{Fe}]
\end{equation}
where $m_p$ is the proton mass, $n_{He} = 0.0977 n_H$
and 
\begin{equation}
n_{Fe} = n_H z_{Fe}/A.
\end{equation}
The electron density is 
\begin{equation}
n_e = n_H + 2 n_{He} + {\bar Z} n_{Fe} = n_H [1.20 + z_{Fe}{\bar Z}/A].
\end{equation}

\section{Radiative Cooling Coefficient}

In the optically thin limit, the cooling coefficient $\Lambda$ 
is defined so that the plasma loses energy by 
radiation at a rate $n_e n_H \Lambda$ erg cm$^{-3}$ s$^{-1}$.
Using the XSPEC program and 
assuming very high iron abundances, 
we have extracted the cooling coefficient $\Lambda_{Fe}(T)$ 
for a pure iron plasma, 
defined so that $n_{Fe}n_e \Lambda_{Fe}(T)$
is the total power radiated per unit volume.
We have also computed the cooling coefficient
$\Lambda_{no Fe}(T)$ for 
an iron-free plasma with solar abundances for 
all the other elements (as listed by Feldman 1992).
These cooling coefficients can be combined to give the 
cooling rate for any iron abundance $z_{Fe}$, 
\begin{equation}
\Lambda(T,z_{Fe}) = 
\Lambda_{no Fe}(T) + (z_{Fe}/A)\Lambda_{Fe}(T).
\end{equation}
As we illustrate in Figure 3,
in an iron-rich plasma the cooling coefficient
$\Lambda(T,z_{Fe})$ can be many orders of magnitude 
greater than that for a plasma with solar abundance.
However, the emission rate $n_e n_H \Lambda(T,z_{Fe})$ 
approaches a finite value as 
$z_{Fe} \rightarrow \infty$ 
that is only about $\sim 280$ times larger than the 
emissivity at solar abundance, 
provided the gas density is held constant.
For $z_{Fe} \gta 100 z_{Fe,\odot}$ 
the peak emission is shifted toward higher 
temperatures.
The cooling coefficients shown in Figure 3 are our analytic 
fits to the XSPEC results for $\Lambda_{no Fe}(T)$
and $\Lambda_{Fe}(T)$. 

\section{Cooling of an Iron-Rich Plasma by Mixing 
and Radiation}

After heating by the reverse shock,
the temperature of the iron core in SNIa remnants 
is about 2-15 times greater than 
that in the surrounding hot interstellar gas (Table 1). 
We shall estimate the subsequent thermal evolution of 
this iron-rich gas as it cools by radiation losses and 
by mixing with the cooler interstellar gas.
The mixing processes that occur when the iron plasma 
encounters the ambient (solar abundance) plasma are 
exceedingly complex.
The Rayleigh-Taylor instability that accompanies the 
deceleration of the stellar remnant greatly increases
the surface area between the iron and ambient plasmas.
This mixing process is further assisted by 
irregularities in the exploding star and by interstellar 
turbulence expected from observations and general
theoretical considerations. 
However, coarse grain mixing of the iron is insufficient
for our purposes. 
Mixing must occur 
on an {\it atomic} level to influence the radiation losses. 
Once the surface area at the plasma interface has been 
enormously increased by coarse grain mixing, 
particle diffusion is essential 
to mix individual iron ions into the ambient gas 
and this diffusion depends on 
the strength and orientation of the local magnetic field.
In view of the profound difficulty in estimating the ion 
mixing time due to diffusion, we shall simply assume 
that this mixing occurs on a timescale $t_m$ that we 
regard as a parameter.

The iron-rich gas cools by two processes: 
(1) thermal mixing with cooler ambient
solar abundance gas and (2) radiative losses.
At any time we assume that a fraction $f$ of the mass of 
colder ambient gas has physically mixed with the current 
iron-rich supernova core. 
Initially when $f = 0$ the gas is essentially pure iron; 
when the mass of mixed ambient gas is much greater than 
that of the SNIa remnant, $f \rightarrow 1$ and the iron 
abundance is nearly that of the ambient gas (with $z_{Fe\odot}$).
We assume for simplicity that this mixing occurs on 
a timescale $t_m$ so that the fraction of cold gas that has 
mixed at time $t$ is given by 
\begin{equation}
f = 1 - e^{-(t/t_m)}.
\end{equation}
Initially at $t = 0$ when $f = 0$ 
we consider a hot plasma at temperature 
$T_h \approx \langle T \rangle_{0.7}$ 
that is sufficiently iron-rich, $z_{Fe,h} \gta z_{Fe,crit}$, 
so that the electrons contributed by iron dominate.
For a plasma of any composition, the ratio of iron mass to 
total mass is $z_{Fe}/(1.39 + z_{Fe})$ (Equation 3). 
Consequently, 
as the iron-rich gas mixes with ambient gas 
at temperature $T_c < T_h$, 
the instantaneous iron abundance $z_{Fe}(t)$ is found from 
\begin{equation}
{z_{Fe} \over {1.39 + z_{Fe}}} =
(1-f) {z_{Fe,h} \over {1.39 + z_{Fe,h}}} 
+ f {z_{Fe\odot} \over {1.39 + z_{Fe\odot}}}. 
\end{equation}

The numerical solution for the thermal variation of the iron-rich 
plasma during a small time interval 
proceeds in two steps: first we determine the 
thermal change due to mixing alone 
then we compute the losses due to radiation during that same time 
interval.
During the mixing step 
the specific thermal energy $\varepsilon = 3P/2\rho = 3kT/2\mu m_p$
varies as the relative mass $f$ of cold gas mixes in.
For a given time step $\delta t = t^{n+1} - t^{n}$ 
Equation (9) can be used to determine 
the change in the fraction of cold gas 
$\delta f = f(t^{n+1}) - f(t^n) \equiv f^{n+1} - f^n$. 
The decreased iron abundance $z_{Fe}^{n+1} = z_{Fe}(f^{n+1})$ 
is found from Equation (10). 
The specific thermal energy varies according to 
\begin{equation}
(1 - f^n + \delta f) {T_m^{n+1} \over \mu^{n+1}} 
= (1 - f^{n}) {T^{n} \over \mu^{n}} 
+ \delta f {T_c \over \mu_c}
\end{equation}
where $\mu^n \equiv \mu(T^{n},z_{Fe}(t^n))$
and $\mu_c \equiv \mu(T_c,z_{Fe\odot})$.
This equation is solved for the intermediate, post-mixed 
temperature $T_m^{n+1}$.

We assume that the entire cooling process occurs at 
constant pressure $P_o$ so 
\begin{displaymath}
P_o = {k T_m^{n+1} \rho^{n+1} \over \mu^{n+1} m_p} 
\end{displaymath}
\begin{equation}
~~~~= {k T_m^{n+1} \rho^{n+1} \over m_p} 
{{ [2.29 + z_{Fe}^{n+1}(1 + {\bar Z})/A]}
\over {[1.39 + z_{Fe}^{n+1}]}}
\end{equation}
can be used to determine the total density $\rho^{n+1}$ 
at this stage of the calculation. 
Intermediate values for 
the number densities $n_H$, $n_{Fe}$ and $n_e$ 
are found from the expressions derived earlier. 
This completes the mixing step of the calculation.

Radiative cooling at constant pressure $P_o$ can be regarded 
as a two-step process corresponding to the two terms 
on the right hand side of the thermal energy equation
\begin{equation}
{ d \varepsilon \over d t} = {P \over \rho^2} { d \rho \over d t}
- {n_e n_H \Lambda(T,z_{Fe}) \over \rho}.
\end{equation}
First we allow the gas to cool from $T_m^{n+1}$ by radiative losses 
at constant density and $z_{Fe}$
\begin{equation}
\delta T = - {T_m^{n+1} \over t_{cool}} \delta t
\end{equation}
where $T_m^{n+1}$ is the intermediate temperature found 
from Equation (11) and 
\begin{equation}
t_{cool}(T,z_{Fe}) = {3 \over 2} { (kT)^2 \over P_o \Lambda} 
{ [2.29 + z_{Fe}(1 + {\bar Z})/A]^2 \over {[1.20 + z_{Fe}{\bar Z}/A]}}
\end{equation}
is the instantaneous radiative cooling time.
However, the temperature at $t^{n+1}$ following the radiative 
energy loss, $T_r^{n+1} = T_m^{n+1} + \delta T$, 
corresponds to a pressure 
$P_r^{n+1} = kT_r^{n+1} \rho^{n+1}/\mu^{n+1} m_p$
that generally is less than the constant pressure $P_o$ assumed.
To correct this, we apply an adiabatic compression corresponding 
to the first term in the thermal energy equation above,
\begin{equation}
T^{n+1} = T_r^{n+1} (P_o/P_r^{n+1})^{2/5}.
\end{equation}
This is the final temperature at time $t^{n+1} = t^n + \delta t$.
To complete this computational cycle, the 
time-advanced densities 
$\rho$, $n_H$, $n_e$ and $n_{Fe}$ are 
determined again from the equation of state (12) 
consistent with constant $P_o$ and the final temperature.

\subsection{A Typical Cooling Evolution}

In Figure 4 we show a typical constant-pressure 
cooling pattern for an iron-rich 
plasma using parameters 
for which cooling by diffusive mixing and radiation losses
are comparable: 
$t_m = 10^7$ yrs, $P_o = 2 \times 10^{-11}$ 
dyn cm$^{-2}$, $z_{Fe,h}(t = 0) = 10^5z_{Fe\odot}$, 
$z_{Fe,c} = z_{Fe\odot}$,
$T_h = 10^8$ K and $T_c = 10^7$ K.
The top panel shows the assumed variation of the fraction 
of cold gas $f(t)$ and the 
iron abundance $z_{Fe}(t)$, both of which vary with time 
as required by Equations (9) and (10).
For the parameters chosen for Figure 4, the gas temperature 
$T(t)$ and the instantaneous cooling time $t_{cool}(t)$ evolve 
in a peculiar manner. 
As the elapsed time continues toward $t_m = 10^7$ yrs,
both the temperature and the cooling
time decrease as expected.
But then $t_{cool}$ suddenly
shoots up by almost two orders of magnitude.
At that time the gas temperature actually increases
toward $T_c = 10^7$ K as
gas at that temperature continues to mix in, overriding
the general decline in temperature expected from radiation losses.
Eventually, at time $\sim 100t_m$ the gas finally cools
by radiation in a catastrophic fashion.
The bottom panel in Figure 4 shows the 
corresponding evolution of several plasma densities.
The proton and electron densities increase monotonically
as gas with solar abundance mixes in.
The density of iron ions $n_{Fe}$ decreases as expected
until the final catastrophic cooling when all densities
increase without limit.
During mixing, 
the electron to iron ion density ratio $n_e/n_{Fe}$ increases
by about $\sim 1000$ as expected from the preceding discussion.
Finally, we note that the change in slope of the cooling
time $t_{cool}$ at $\log T = 7.4$ is caused by the slope
change at this temperature in $\Lambda(T)$ (Figure 3).

In Figure 5 we plot the final cooling times 
(to $T = 10^5$ K) for four 
assumed initial temperatures for the iron-rich plasma 
$T_h$ and four pressures $P_o$, corresponding to different 
radii in the the hot gas atmospheres of luminous elliptical galaxies.
For each calculation the initial iron abundance in the 
supernova ejecta is $z_{Fe,h} = 10^5z_{Fe\odot} = 183.$
The mixing times $t_m$ are varied over a wide range, 
recognizing the uncertainty in this important parameter. 
For all calculations the ambient interstellar gas has 
temperature $T_c = 10^7$ K and abundance $z_{Fe,c} = z_{Fe\odot}$. 
The constant pressure cooling times of the (unmixed) 
interstellar gas 
are shown as horizontal dashed lines for each 
assumed pressure $P_o$.

As cold gas is mixed 
into the radiating iron-rich gas on progressively longer 
timescales $t_m$, the cooling times plotted in 
Figure 5 first increase, then sharply 
decrease, then increase again.
Similar non-monotonic 
$t_{cool}(t_m)$ profiles occur if $T_c$ or $z_{Fe,c}$ are changed.
In the following we explain this counterintuitive behavior 
in more detail.

When $T_h = T_c$, as shown in the upper left panel of Figure 5,
the mixing does not cool the iron-rich gas but only dilutes 
it with gas having solar abundance. 
As the dilution or mixing time $t_m$ increases smoothly, 
the cooling time
$t_{cool}$ suddenly drops precipitously at a particular 
value of $t_m = t_{m,crit}$. 
When $t_m \ll  t_{m,crit}$, the plasmas with different 
abundance tend to mix completely before radiative cooling occurs 
and $t_{cool}$ 
is just the cooling time of the 
interstellar gas with solar abundance (dashed lines for 
each $P_o$). 
Eventually, when $t_m$ just exceeds $t_{m,crit}$, 
the supernova gas cools rapidly when it still has 
an iron abundance much greater than solar.
As can be seen in Figure 5, the transition in $t_{cool}(t_m)$ is 
very sharp indeed -- the cooling time drops by $\sim 1000$.

In the other three panels of Figure 5 we illustrate the 
cooling time variation $t_{cool}(t_m)$ 
for the iron rich supernova ejecta 
using more realistic initial temperatures $T_h > T_c = 10^7$ K.
As $t_m$ approaches $t_{m,crit}$ from below, 
the cooling time increases. 
This can be understood by examining the cooling evolution 
$T(t)$ in more detail.
The temperature and cooling time minima
(as in Figure 4) 
become more pronounced when $T_h > T_c$, 
but after this minimum the gas temperature returns to $T_c$
and cools on a longer timescale.
When $T_h > T_c$, the sharp decrease in the cooling time 
in Figure 5 
at $t_{m,crit}$ is followed by progressively longer 
cooling times as $t_m$ increases further.

To demonstrate further the curious non-monotonic relationship 
between the cooling and mixing times, we 
illustrate in Figure 6 the cooling evolution for 
four values of $t_m = 10^5$, $10^7$, $2 \times 10^7$ 
and $2 \times 10^8$ years. 
All four cooling calculations refer to pressure 
$P_o = 2 \times 10^{-11}$ dyn cm$^{-2}$ and $T_h = 10^8$ K 
(as in the lower left panel of Figure 5).
In the central panel of Figure 6 it is seen that the cooling time 
increases, then sharply decreases, then increases again 
as $t_m$ increases monotonically. 
>From the upper panel of Figure 6, 
if $t_m = 10^5$ and $10^7$ years, the gas mixes 
completely to interstellar values before radiatively cooling, 
and the well-mixed iron cools radiatively on a timescale 
similar to that of the interstellar gas (with $z_{Fe\odot}$).
However, for $t_m = 2 \times 10^7$ or $2 \times 10^8$ years, 
the supernova enriched gas radiatively cools 
before it fully mixes with the ambient gas 
and the supernova ejecta is selectively cooled.
The central panel of Figure 6 
shows how early mixing can prolong 
the final cooling time to low temperatures. 
For these values of $P_o$ and $T_h$, the most rapid cooling 
occurs for $t_m \approx 2 \times 10^7$ years when 
the iron-rich gas cools about 17 times faster than the 
ambient gas.
The bottom panel of Figure 6 shows the variation of the 
instantaneous cooling time $t_{cool}(t)$ evaluated with 
Equation (15). 
If the gas mixes before it cools significantly by radiative 
losses, the cooling time passes through a minimum and 
rises sharply thereafter, greatly prolonging the final 
rapid cooling that eventually occurs.

The relationships between the cooling and mixing
times shown in each panel of 
Figure 5 appear to have a self-similar variation with each 
pressure $P_o$.
In Figure 7 
the variations of $t_{cool}$ with $t_m$ in Figure 5 
are plotted again with both times normalized with the 
cooling time of the ambient gas 
$t_{cool,c} = t_{cool}(T_c,P_o,z_{Fe\odot})$.
In Figure 7
all curves in each panel of Figure 5 merge into a single curve 
valid for all pressures $P_o$, confirming the self-similarity.
 
The self-similarity in Figure 7 is surprising 
since the instantaneous cooling time expressed by Equation (15)
depends on $z_{Fe}(t)$ which evolves differently in each 
cooling calculation with different $P_o$.
However, as we illustrate in Figure 6, the gas either
cools when completely mixed or it cools when 
$z_{Fe} \gta 100z_{Fe,c}$ (where $z_{Fe,c} = z_{Fe\odot}$)
-- there are no intermediate cases. 
In either limit, as $z_{Fe} \rightarrow z_{Fe\odot}$ or 
$z_{Fe} \rightarrow \infty$, $t_{cool}$ approaches a constant 
value independent of $z_{Fe}$, 
and this is the origin of the self-similarity in Figure 7.

The magnitude of the sharp transitions in
$t_{cool}$ at $t_{m,crit}$ in Figure 7 can be understood as
the ratio of cooling times between the interstellar gas
($T_c$) and the iron rich ejecta ($T_h$).
As $t_m/t_{cool,c}$ increases through the discontinuity,
the cooling time of the gas drops by a factor
\begin{displaymath}
{ t_{cool}(T_h) \over t_{cool}(T_c)}
\approx \left( { T_h \over T_c} \right)^2
\left[
{ \Lambda_{noFe}(T_c) \over \Lambda_{Fe}(T_h)}
{ 1.20 \over (2.29)^2 }
{ (1 + {\bar Z})^2 \over {\bar Z}} 
\right].
\end{displaymath}
The quantity in square brackets is $\sim 10^{-3}$ 
when $T_h = T_c$, but increases 
approximately as $\sim T_h/T_c$ when $T_h > T_c$.
For $T_h \gta 3 \times 10^8$ K
the cooling time of the iron-rich gas can exceed that
of the ambient gas at $T_c = 10^7$ K if $t_m$ is
sufficiently long.
In general the critical mixing time $t_{m,crit}$ 
is proportional to $T_h/P_o$.

\section{Iron Diffusion in NGC 5044}

The discussion above suggests that rather small variations 
of uncertain parameters -- $t_m$, $E_{sn}$, $T_h$, etc. -- determine 
whether iron produced in SNIa explosions 
in elliptical galaxies either cools much 
faster than the ambient gas or thermally mixes with it 
and cools on a much longer timescale. 
For typical mean post-shock temperatures of SNIa iron in Table 1,
$T_h \approx 7\times 10^7$ K, 
the radiative cooling time is about an order of magnitude 
less than that of the ambient gas $T_c \approx 10^7$ K 
provided the mixing time is larger than a critical value,
\begin{displaymath}
t_{m,crit} \approx  2.54 \times 10^{7} 
\left( { P_o \over 10^{-11}~{\rm dyn~cm}^{-2}} \right)^{-1}
\end{displaymath}
\begin{equation}
~~~~~~~~~~~\times 
\left( {T_h \over 7 \times 10^7~{\rm K}} \right)~~~{\rm yrs}
\end{equation}
where the coefficient is found from Figure 5.
Guided by Figures 6 and 7, to avoid rapid radiative cooling,
the iron-rich SNIa gas must mix
in time $t_m < t_{m,crit}$ with enough interstellar gas
to reduce the iron abundance
to at least $z_{Fe,crit} \sim 100z_{Fe,\odot}$.
The mass of interstellar gas that 
satisfies this condition is 
\begin{equation}
M_{c,mix} = {1.4 M_{Fe,sn} \over 100 z_{Fe\odot}}
= 5.3 { (z_{Fe,crit}/z_{Fe\odot}) \over 100} ~~~ M_{\odot}~~~
\end{equation}
where  $M_{Fe,sn} = 0.7$ $M_{\odot}$ is the mass of iron in each
SNIa remnant and 1.4 accounts for the mass in helium.

The question of interest is whether the iron-rich remnant
can thermally mix with $M_{c,mix}$ of ambient gas in time
$t_m < t_{m,crit}$.
To avoid premature rapid radiative cooling of iron in SNIa
remnants,
the mixing must be totally complete on the
{\it particle} level; mixing of small but finite fluid masses
will not alter the radiative cooling rates.
(Note that mixing on particle scales
does not necessarily produce
a spatially uniform final iron abundance on galactic scales.)
Unfortunately, the complexity of the irregular morphology of
the SNIa iron and uncertainties about the rate of
physical diffusion preclude a rigorous calculation
of the mixing time $t_m$.
Nevertheless, it is easy to estimate the minimum mean
free path $\lambda_{diff}$ required for iron ions
to mix with mass $M_{c,mix}$ in $t_m < t_{m,crit}$, 
and this can be compared with
the plasma mean free path and Larmor radius.
We make these estimates using the well-known
hot gas atmosphere in the luminous
elliptical galaxy NGC 5044.
Observations suggest that a large fraction, but possibly
not all, of the SNIa iron produced in NGC 5044
may have indeed thermally mixed into the hot interstellar gas 
(Mathews et al. 2004).
The gas temperature $T_c$, electron density $n_{ec}$, 
gas density $\rho_c$,
and stellar density $\rho_*$ in NGC 5044
are listed in Table 2 at four representative galactic radii. 
For comparison the effective radius of NGC 5044 is $R_e = 10.0$ kpc.
Local values of the critical mixing time $t_{m,crit}$ are 
also listed at each radius.

We wish to estimate the diffusion of iron into the ambient 
hot gas containing mass $M_{c,mix}$ around the stellar remnant. 
For simplicity we ignore the outermost stellar 
layers containing Si and other elements and the complex geometrical 
disturbances caused by the Rayleigh-Taylor instability. 
We suppose instead that a spherical, pure-iron remnant 
of radius $r_* \approx 19$ pc  
and mass $M_{sn} = 1.4$ $M_{\odot}$ 
is in direct contact with ambient gas of abundance $z_{Fe\odot}$ 
when the diffusive mixing begins. 
It is seen from Figure 1 that the temperature in this outer 
region drops sharply with radius,
$T(r_{pc}) \approx 1.35\times 10^5 T_c r_{pc}^{-3.15}$ K 
for $r_* = 19 < r < 40$ pc where 
$T_c$ is the initial gas temperature prior to the explosion. 
We assume that this radial dependence 
of the post-shock equilibrium temperature in Figure 1 can be 
scaled for other values of $T_c$. 
The gas density in this region is given by 
$\rho(r_{pc}) \approx 1.14 \times 10^{-29} n_{ec} (P/P_s)
r_{pc}^{3.15}$ gm cm$^{-3}$ where $P_s = 2.65 \times 10^{-11}$ 
dy cm$^{-2}$ is a reference pressure.
The radius $r_{mix}$ containing the critical 
ambient mixing mass $M_{c,mix}$, shown in Table 2, 
can be found by integrating the mass outwards from 
the contact discontinuity $r_{*}$ at the stellar-interstellar 
interface.
The mass-weighted mean temperature ${\bar T}$ 
and mean density ${\bar n}_e$ in the shell 
$r_* < r < r_{mix}$ is found from a similar integration.
Values of these quantities at each galactic radius 
are listed in Table 2.

The distance $r_{diff}$ that iron ions diffuse 
in time $t_{m,crit}$ is $r_{diff} \sim (D t_{m,crit})^{1/2}$
where the diffusion coefficient
$D \sim \langle v_{Fe} \rangle \lambda_{diff}$ depends on the
(Maxwellian) mean thermal velocity of iron ions 
$\langle v_{Fe}\rangle = (8 k {\bar T} / \pi 56 m_p)^{1/2}$
and $\lambda_{diff}$ is their effective mean free path. 
The condition that iron ions diffuse a distance 
$r_{diff} \approx r_{mix} - r_*$ 
in the critical cooling time $t_{m,crit}$ 
can be used to determine the minimum 
iron mean free path,
$\lambda_{diff} \approx (r_{mix} - r_*)^2 / 
\langle v_{Fe} \rangle t_{m,crit}$ to avoid radiative cooling.
Values of $t_{m,crit}$ (calculated with $T_h = 7.5\times 10^7$ K)
and $\lambda_{diff}$ 
are listed in Table 2 at each selected radius in NGC 5044.

The likelihood that the SNIa iron mixes in 
less than time $t_{m,crit}$ can be evaluated by comparing 
the effective mean free path for diffusion $\lambda_{diff}$ 
with the field-free Spitzer plasma mean free path 
and the Larmor radius of the iron ions. 
The crossection for Coulomb scattering of iron ions in 
an unmagnetized plasma is 
\begin{displaymath}
\sigma_C = 2 \pi^{1/2} e^4 (kT)^{-2} Z_{Fe}^2 Z_H^2 \ln \Lambda
\end{displaymath}
\begin{displaymath}
~~~~~~~~~\approx 1.74 \times 10^{-15} (T/10^7~{\rm K})^{-2}~~~{\rm cm}^2,
\end{displaymath}
where we adopt $\ln \Lambda = 40$
(e.g. Chuzhoy \& Loeb 2004). 
The iron ions are assumed to diffuse through 
protons, ignoring helium ions for simplicity, 
and for $z_{Fe} \sim 100 z_{Fe,\odot}$
$n_H \gg n_{Fe}$. 
The field-free plasma mean free path for the diffusing 
iron is then 
$\lambda_{spitz} = 1 / {\bar n}_H \sigma_C\,=\,
0.024 (T/10^7\,{\rm K})^2\,(n_e/0.01\,{\rm cm}^{-3})^{-1}$ 
pc, where 
${\bar n}_{H} = {\bar n}_{e}(4-3\mu)/(2+\mu)$ 
is the mean proton density in the diffusion shell.
Comparing $\lambda_{spitz}$ listed in 
Table 2 with the smaller mean free path $\lambda_{diff}$ 
required to avoid rapid iron-rich cooling, 
we conclude that $\lambda_{spitz}$ is generally large enough 
to ensure that the SNIa iron mixes into the ambient gas 
in $t_m \lta t_{m,crit}$. 
Mixing is particularly likely within the central 
galaxy $r \lta R_e$. 
But mixing is only guaranteed if the iron  
diffusion is unaffected by the local magnetic field. 

If magnetic fields are present, the ability of the SNIa iron to 
mix may be lessened if the Larmor radius $r_{L,Fe}$ is 
less than $\lambda_{spitz}$.
Values of the Larmor radius of iron
ions, $r_{L,Fe} = 56 m_p \langle v_{Fe} \rangle c /{\bar Z} e B$, 
listed in Table 2 are evaluated with
${\bar Z} = 21$ and $B = 3 \times 10^{-6}$ gauss.
Clearly $r_{L,Fe} \ll \lambda_{spitz}$.
If the magnetic fields are coherent over scales much larger than 
$r_{L,Fe} \sim 10^9$ cm, the diffusive mixing of iron 
will be greatly reduced except along field lines. 
>From Figure 1 we see that the temperature is higher and the 
gas density is lower in the diffusion shell $r_{mix} - r_*$ 
compared to $T_c$ and $\rho_c$ in the undisturbed, pre-explosion gas.
The radial expansion that lowered the density may have stretched and 
oriented the magnetic field in a more radial direction, 
enhancing the diffusive mixing. 
Alternatively, if the magnetic field is highly 
tangled on a range of small scales, 
as in strong MHD turbulence, the effective plasma mean free path 
may not be much less than $\lambda_{spitz}$ 
(Narayan \& Medvedev 2001). 
MHD turbulence may be necessary for the efficient 
mixing of SNIa iron into the hot gas. 

One possible argument against the selective cooling of iron-rich
SNIa remnants
is that there are no known elliptical galaxies 
with stellar iron abundances greater than 1-2$z_{Fe\odot}$.
However, such high stellar iron abundances 
would not be expected in elliptical galaxies 
since the number of these stars and their contribution to the 
integrated spectrum would be very small.
Even if all the super-iron-rich 
gas in SNIa remnants cooled, the total star formation 
rate in NGC 5044 would be only 
$M_{c,mix} \alpha_{sn} (M_{*t}/M_{sn}) \approx 0.04$ 
$M_{\odot}$ yr$^{-1}$. 
The total mass of these stars would be an unobservable 
fraction of the stellar mass of NGC 5044, $M_{*t} = 3.4 \times
10^{11}$ $M_{\odot}$.

\section{SNIa Iron Inhomogeneities}

The iron deposited in the hot interstellar gas by SNIa explosions 
is inherently inhomogeneous.
The amount of inhomogeneity can be estimated 
by comparing the mean distance 
between SNIa events after some characteristic time 
to the distance that iron has diffused during that time.
The rate that SNIa supernova remnants occur in elliptical 
galaxies is
$\alpha_{sn} \rho_* / M_{sn}$ SNIae cm$^{-3}$ s$^{-1}$ 
where $\rho_*$ is the stellar density and 
\begin{displaymath}
\alpha_{sn} = 3.17 \times 10^{-20} {\rm SNu}
{(M_{sn}/M_{\odot}) \over \Upsilon_B} \approx 1.01 \times 10^{-21} ~~~{\rm
s}^{-1}.
\end{displaymath}
SNu = 0.16 is the current E-galaxy supernova rate 
in units of SNIae per 
$10^{10} L_B$ in 100 years (Cappellaro et al. 1999),
$M_{sn} = 1.4$ $M_{\odot}$ is the typical mass ejected and 
we adopt 
$\Upsilon_B = 7$ for the stellar mass to B-band light ratio.
After time t the mean distance between the centers of SNIa remnants is 
\begin{displaymath}
d_{sn} = \left( { M_{sn} \over t \alpha_{sn} \rho_*}
\right)^{1/3}.
\end{displaymath}
The local stellar density $\rho_*(r)$ 
in NGC 5044 shown in Table 2 is estimated 
from a de Vaucouleurs profile with total stellar mass 
$M_{*t} = \Upsilon_B L_B = 3.4 \times 10^{11}$ $M_{\odot}$
and effective radius $R_e = 10.0$ kpc.

In Table 2 we list values of $d_{sn}$ based on 
$t = t_{cool}(r)$, 
the constant-pressure radiative cooling time for 
interstellar gas with $T = 10^7$ K and $z_{Fe} = z_{Fe\odot}$.
The time $t_{cool}$ is a conservative measure of the radial 
flow time in the local interstellar gas.
This time must be compared with 
the time for iron ions to diffuse a distance $d_{sn}$
at the Spitzer rate, 
$t_{diff} \approx d_{sn}^2/\langle v_{Fe} \rangle \lambda_{spitz}$. 
Comparing these times listed in Table 2,
we see that $t_{diff} \lta t_{cool}$ for 
$r \lta R_e$. 
But at $r \gta R_e$, the iron may not diffuse fast enough 
to achieve uniform abundance in the hot gas.
It is likely, however, that the iron can be mixed to $d_{sn}$ 
more efficiently by turbulent diffusion in the 
hot gas than our estimate suggests. 

It is interesting to explore how the iron emission depends on the
degree of iron inhomogeneity.
Consider for simplicity a fixed volume $V$
containing a mass $M$ of gas
with uniform solar iron abundance.
Then imagine that the iron occupies a decreasing fraction
$f_z = V_{Fe}/V \le 1$ of the volume
but that the pressure, the density
and the total iron mass remain constant.
Provided $z_{Fe\odot} < z_{Fe} < z_{Fe,crit}$,
the bolometric iron luminosity
\begin{displaymath}
L_{Fe} = n_{Fe} n_e \Lambda_{Fe}(T) V_{Fe}
\end{displaymath}
remains constant with decreasing $f_z \sim z_{Fe\odot}/z_{Fe}$
because of the conservation of iron
$n_{Fe} V_{Fe}$ and $n_e$ and $\mu = 0.61$ are unchanged
as long as most of the electrons come from H and He.
The ratio of total energy radiated
when the iron occupies a volume fraction $f_z$
to the energy radiated
by uniformly distributed iron is
$E_{Fe}/E_{Fe,0} \sim L_{Fe} t_{cool} / L_{Fe,0} t_{cool,0}$
and decreases in pace with
\begin{displaymath}
{t_{cool}\over t_{cool,0}} \sim
{\Lambda_{noFe}/ \Lambda_{Fe} + z_{Fe\odot}/ A \over
\Lambda_{noFe}/\Lambda_{Fe} + z_{Fe}/A }.
\end{displaymath}
As $z_{Fe}$ approaches $z_{Fe,crit}$
the ratio $L_{Fe}/L_{Fe,0}$ decreases slightly and
asymptotically reaches the value
$1.39 {\bar Z}/1.2 A \sim 0.44$.
The ratio of total energy radiated
by the iron approaches
\begin{displaymath}
{E_{Fe}\over E_{Fe,0}} \sim
0.62 {L_{Fe}\over L_{Fe,0}}
\left[ {\Lambda_{noFe}\over \Lambda_{Fe}}
+ {z_{Fe\odot}\over A} \right]
{A^2\over {\bar Z}} \sim 0.02.
\end{displaymath}
So far we have
assumed that the iron-rich region is in buoyant
equilibrium with the surrounding gas (i.e. it has the same density).
If instead we assume that the iron-rich gas is in thermal
equilibrium with the initial ambient gas,
$L_{Fe}/L_{Fe,0}$ and $E_{Fe}/E_{Fe,0}$ behave as in the
preceding case
when $z_{Fe}<z_{Fe,crit}$,
but when $z_{Fe}\ge z_{Fe,crit}$
$L_{Fe}/L_{Fe,0}$ increases  to
$2.29 {\bar Z}/1.2 (1+{\bar Z}) \sim 1.8$
and
\begin{displaymath}
{E_{Fe} \over E_{Fe,0} } \sim
0.23 {L_{Fe}\over L_{Fe,0}}
\left[ {\Lambda_{noFe}\over \Lambda_{Fe}}
+ {z_{Fe\odot}\over A} \right]
{(1+{\bar Z})^2\over {\bar Z}} \sim 0.0036.
\end{displaymath}

Morris \& Fabian (2003) discuss the
radiative cooling of iron-rich regions
(with $0 < z_{Fe} < z_{Fe,crit}$)
that have the same initial temperature and pressure
as the surrounding gas.
In this case, small iron-rich
regions with high $z_{Fe}$ cool rapidly, but
radiate a total thermal energy that is
$\sim f_z = V_{Fe}/V$
times less than that of the surrounding plasma.
When integrated over time,
this type of selective transient cooling of iron-rich regions emits
less iron emission than if the same mass of iron cooled
in a gas with uniform abundance
(Morris \& Fabian 2003).

\section{Conclusions}

We have explored the possibility that 
some of the iron produced in SNIa explosions in 
elliptical 
galaxies may cool much faster than ambient gas having 
a normal hot gas abundance  
$\sim z_{Fe\odot}$. 
The temperature evolution of the iron-rich gas is 
governed by optically thin radiation loss and 
by diffusive mixing with the ambient interstellar gas.  
For simplicity we have ignored any differential radial 
motion of the SNIa remnants relative to the surrounding gas.
We assume that a fraction by mass $f(t) = 1 - e^{-(t/t_m)}$ 
of hot interstellar or group gas physically mixes with the SNIa iron, 
regarding $t_m$ as an unknown mixing time. 
Radiative cooling and mixing are assumed to occur at constant pressure.

\vskip.1in
\noindent
Our main conclusion is:

\vskip.1in
\noindent
(1) The question whether 
iron in SNIa remnants can successfully diffuse into 
the hot interstellar gas in X-ray luminous elliptical 
galaxies depends on the uncertain magnetic field 
strength and geometry. 
If there is no field, the iron is expected to mix without 
catastrophic radiative cooling. 
However, if the magnetic field reduces the 
plasma mean free path of the iron ions by more than 
$\sim 30$ relative to the field-free value, 
the iron is likely to cool with very few 
observational consequences.  

\vskip.1in
\noindent
This conclusion is supported by the following 
more detailed results: 

\vskip.1in
\noindent
(2) Iron cores of SNIa remnants that have reached hydrostatic 
equilibrium with ambient gas at $T_c \approx 10^7$ K 
in elliptical galaxies typically have 
mass-weighted temperatures $T_h \approx 7T_c$. 

\vskip.1in
\noindent
(3) Iron-rich plasmas with abundances 
$30z_{Fe\odot} \lta z_{Fe} \lta z_{Fe,crit}$ 
cool rapidly by optically thin (iron) radiation losses
with emissivity 
$\epsilon_{Fe} \approx 1.20 n_H^2  (z_{Fe}/A)
\Lambda_{Fe}(T)$ erg s$^{-1}$ cm$^{-3}$ 
where $A = 56$ is the atomic weight of iron.
The critical iron abundance 
$z_{Fe,crit} = 1.2A/{\bar Z} \approx 3.2 \sim 1750 z_{Fe\odot}$
is the abundance at which iron contributes the 
same number of electrons as H plus He, 
both assumed to be fully ionized, 
evaluated with ${\bar Z} = 21$.
When $z_{Fe} > z_{Fe,crit}$ the emissivity of the iron-rich 
gas asymptotically approaches a constant value 
$\sim 280$ times larger than the emissivity of 
solar abundance gas at the same density.

\vskip.1in
\noindent
(4) If the time $t_m$ required for the iron-rich SNIa core 
to diffusively mix with ambient 
interstellar gas at constant pressure 
$P$ exceeds a critical value 
$t _{m,crit} \sim 2.5 \times 10^7 (P/(10^{-11}~{\rm dy~cm}^{-2}) 
(T_h/7\times 10^7~{\rm K})$ yrs, 
the cooling time of the iron-rich gas is 
reduced by $\sim 10^{-3} (T_h/T_c)^2$, 
where $T_h$ and $T_c$ are the temperatures of the 
iron in a young supernova remnant and the hot ambient 
gas respectively.
This mixing must occur among iron and hydrogen ions, 
not on small but finite mass scales. 

\vskip.1in
\noindent
(5) The  minimum effective mean free path of diffusing iron ions
$\lambda_{diff}$ required to successfully mix the iron-rich SNIa 
remnant in time $t_m < t_{m,crit}$ is 
$\gta 30$ times smaller than the 
plasma mean free path $\lambda_{spitz}$, 
supporting the possibility that the SNIa iron successfully mixes 
with the surrounding gas.
However, the Larmor radius of iron ions in microgauss fields 
is very much smaller, 
$r_{L,Fe} \sim 10^{-10} \lambda_{spitz}$. 
Effective mixing of SNIa iron into the surrounding gas 
may be possible only if the magnetic field is highly tangled on 
small scales as in strong MHD turbulence or is 
stretched in the radial direction.


\vskip.1in
\noindent
(6) Even if all the iron produced 
by SNIae in elliptical galaxies formed into super-iron-rich stars, 
the number of these 
stars would be insignificant and unobservable. 

\vskip.1in
\noindent
(7) If the iron diffuses at or near the Spitzer rate, 
the SNIa iron in elliptical galaxies may be 
mildly inhomogeneous, especially for $r \gta R_e$.

\vskip.1in

As we discussed in the Introduction, 
SNIae are thought to furnish $\sim$ 3/4 
of the iron in the hot gas of massive clusters. 
Most of the iron is thought to be produced
at early times,
as indicated by the approximately constant iron abundance in
the hot gas of rich
galaxy clusters out to redshift $z \sim 1.2$ (Tozzi et al. 2003).
This implies a much higher SNIa rate at $z\gta 1$ than at the
present time.
X-ray luminous groups like NGC 5044, that appear to be 
gravitationally closed systems, contain less than half the iron 
found in rich clusters (Buote, Brighenti \& Mathews 2004).
The origin of this discrepancy is not understood.
If iron-poor groups like NGC 5044 experienced the same enrichment
history as rich clusters, 
it follows that much of the iron in groups has been
buoyantly transported to unobservably large radii. 
Group winds may occur, but not in luminous groups like 
NGC 5044 that are baryonically closed.

If the metals are preferentially expelled on smaller galactic scales, 
this separation may extend to group scales in a natural way. 
In support of this idea, we speculate that that galaxies 
soon after formation are
surrounded by low density halos of iron-rich, high-entropy gas.
These halos could be produced by SNII and SNIa-driven 
winds, by the energy of low level AGN activity in elliptical
galaxies,
as we have discussed in Mathews et al. (2004) and elsewhere, 
or possibly by the buoyancy of individual SNIa remnants, 
particularly in dwarf galaxies.
Later, when these galaxies enter the virial radius of a group 
and confront the relatively denser gas in the group, 
their tenuous iron-rich halos would be stripped.
As more gas accumulates in the group, 
these high-entropy halos may buoyantly flow toward larger radii 
(and entropy) in the group gas
at some fraction of the sound speed.
Galactic halo bubbles of this type moving at $\sim 200$ km s$^{-1}$  
(about half the group gas sound speed) may move out 300 kpc, 
beyond the outermost gas currently observable in NGC 5044
and other similar iron-poor groups (Buote et al. 2004) 
in only $\sim 1.5$ Gyrs.
Even if these hypothetical iron-rich, high-entropy, 
low-density halo bubbles moved more 
slowly, they would be difficult to observe in the outer 
regions of groups because of their relatively small size 
and low X-ray emissivity.
Finally, the iron-rich gas in these galaxy halo bubbles 
may radiatively cool in $\sim 10^9$ yrs 
and be removed from observation.

On smaller scales within group and cluster-centered 
elliptical galaxies, it is likely 
that the hot iron-rich gas within individual SNIa remnants is 
buoyantly transported several kpcs to larger radii.
This may explain the 
central abundance minima observed in NGC 5044 and 
other similar groups and clusters.

If buoyant transport of iron-rich galactic halos in groups 
are the explanation for their apparent iron deficiency, 
groups like NGC 5044 should
be surrounded by (as yet undetected) halos 
of very tenuous hot gas with 
an unusually high iron abundance.
Using the observational data for NGC 5044 
from Buote et al. (2004), 
hot gas detected within $r < 370$ kpc 
has a low mean gas phase iron abundance, 
$\langle z_{Fe} \rangle_{in} \approx 0.21 z_{Fe\odot}$. 
This inner region of NGC 5044 contains a total 
mass of $12.6 \times 10^{11}$ $M_{\odot}$ of hot gas. 
Buote et al. (2004) estimate that 
the total gas mass extrapolated to the virial
radius $r_{vir} = 870$ kpc is $M_{b,tot} = 45.5 \times 10^{11}$ 
$M_{\odot}$ 
and this is approximately consistent with the cosmic baryon ratio. 
If the overall total gas phase iron abundance in NGC 5044 is the 
same as that of rich clusters, 
$\langle z_{Fe} \rangle_{tot} \approx 0.4 z_{Fe\odot}$, 
the unobserved hot gas (of mass $\sim 32.9 \times 10^{11}$ 
$M_{\odot}$) 
beyond 370 kpc associated with the virialized dark halo of 
NGC 5044 would need to have an average iron 
abundance of $\langle z_{Fe} \rangle_{out} \approx 0.5 z_{Fe\odot}$
whether or not this hot gas is gravitationally bound to NGC 5044. 
Much of this as yet unobserved hot gas 
probably has an even higher iron 
abundance since the most distant 
iron abundance observed at $r = 300$ kpc, 
$z_{Fe} \approx 0.15z_{Fe\odot}$, is very low. 
If iron is radially segregated this way in groups that 
later merge to form clusters, 
then we would expect pronounced iron inhomogeneities in rich clusters 
on scales of $\sim 10^{12}$ $M_{\odot}$.

If future more detailed gasdynamic models 
can explain the group-cluster iron discrepancy and the central iron 
abundance dips in a natural way, there may be no compelling 
reason to assume that SNIa iron is lost by radiative cooling, 
but at present this type of remnant cooling 
must remain a possibility. 

\vskip.4in
Studies of the evolution of hot gas in elliptical galaxies
at UC Santa Cruz are supported by
NASA grants NAG 5-8409 \& ATP02-0122-0079 and NSF grants
AST-9802994 \& AST-0098351 for which we are very grateful.





\clearpage

\makeatletter
\def\jnl@aj{AJ}
\ifx\revtex@jnl\jnl@aj\let\tablebreak=\nl\fi
\makeatother

\begin{deluxetable}{lccc}
\tablewidth{0pc}
\tablecaption{MEAN SNIA IRON REMNANT TEMPERATURES\tablenotemark{a}}
\tablehead{
\colhead{$n_e$} &
\colhead{$T$} &
\colhead{$\langle T \rangle_{0.35}$} &
\colhead{$\langle T \rangle_{0.70}$} \\
\colhead{(cm$^{-3}$)} &
\colhead{(K)} &
\colhead{(K)} &
\colhead{(K)}}
\startdata
0.01  &   $1.0 \times 10^7$ & $8.1 \times 10^7$ & $6.7 \times 10^7$
\cr
0.03  &   $1.0 \times 10^7$ & $7.8 \times 10^7$ & $6.6 \times 10^7$
\cr
0.003  &   $1.0 \times 10^7$ & $7.7 \times 10^7$ & $6.6 \times 10^7$
\cr
0.001  &   $4.0 \times 10^6$ & $7.0 \times 10^7$ & $5.8 \times 10^7$
\cr
0.001  &   $1.0 \times 10^8$ & $1.9 \times 10^8$ & $1.6 \times 10^8$
\cr
\enddata
\tablenotetext{a}{$\langle T \rangle_{0.35}$ and $\langle T
  \rangle_{0.70}$ are the mean mass-weighted 
final temperatures of iron in SNIa
remnants within 0.35 and 0.70 $M_{\odot}$ respectively.
Remnants
explode into uniform gas with density $n_e$ and temperature $T$.
For all calculations $t_i = 1 \times 10^7$ s,
$E_{sn} = 10^{51}$ ergs and $M_{sn} = 1.4$ $M_{\odot}$.}
\end{deluxetable}

\clearpage

\begin{deluxetable}{lccccc}
\tablewidth{0pc}
\tablecaption{IRON DIFFUSION IN NGC 5044}
\tablehead{
}
\startdata
$r$ & (kpc) & 1.00 & 3.16 & 10.00 & 31.6 \cr
$T_c$ & (K)                    & $7.22 \times 10^{6} $ & $7.86 \times 10^{ 6}$ & $8.98 \times 10^{6 }$ & $1.32 \times 10^{7} $ \cr
$n_{ec}$ & (cm$^{-3}$)      	& $8.47 \times 10^{-2}$ & $3.88 \times 10^{-2}$ & $1.48 \times 10^{-2}$ & $4.15 \times 10^{-3}$ \cr
$\rho_c$&(g cm$^{-3}$)   		& $1.64 \times 10^{-25}$ & $7.53 \times 10^{-26}$ & $2.86 \times 10^{-26}$ & $8.05 \times 10^{-27}$ \cr
$P_c$ & (dy cm$^{-2}$)            & $1.60 \times 10^{-10}$ & $8.00 \times 10^{-11}$ & $3.48 \times 10^{-11}$ & $1.44 \times 10^{-11}$ \cr
$\rho_*$&(g cm$^{-3})$ 		& $1.19 \times 10^{-22}$ & $1.06 \times 10^{-23}$ & $5.81 \times 10^{-25}$ & $1.69 \times 10^{-26}$ \cr
$t_{m,crit}$ & (yrs)		& $1.70 \times 10^{6}$ & $3.40 \times 10^{6}$ & $7.82 \times 10^{6}$ & $1.89 \times 10^{7}$ \cr
$r_{mix}$ & (pc) 		& 19.9 & 22.3 & 28.4 & 39.9 \cr
${\bar T}$ & (K) & $8.72\times10^7$ & $7.65\times10^7$ & $5.13\times10^7$ & $3.04\times10^7$ \cr
${\bar n}_e$ & (cm$^{-3}$)      & $6.90 \times 10^{-3}$ & $3.93 \times10^{-3}$ & $2.55 \times 10^{-3}$ & $1.78 \times 10^{-3}$ \cr
$\langle v_{Fe} \rangle$ & (km s$^{-1}$)& 181 & 170 & 139 & 107 \cr
$\lambda_{diff}$ & (pc)	        & 0.00431 & 0.0222 & 0.0836 & 0.215 \cr
$\lambda_{spitz}$ & (pc)        & 2.46 & 3.32 & 2.31 & 1.16 \cr
$r_{L,Fe}$ & (pc)               & $5.43 \times 10^{-10}$ & $5.09 \times 10^{-10}$ & $4.17 \times 10^{-10}$ & $3.21 \times 10^{-10}$ \cr
$t_{cool}$ & (yrs)            & $6.9\times10^7$ & $1.4\times10^8$ & $3.2\times10^8$ & $7.7\times10^8$ \cr
$d_{sn}$ & (pc)               & 72 & 127 & 253 & 614 \cr 
$t_{diff}$ & (yrs)            &$1.12\times10^{7}$&$2.77\times10^{7}$&$1.95\times10^{8}$&$2.96\times10^{9}$\cr
\enddata
\end{deluxetable}

\clearpage
\begin{figure}
\includegraphics[bb=90 216 522 569,angle= 0]
{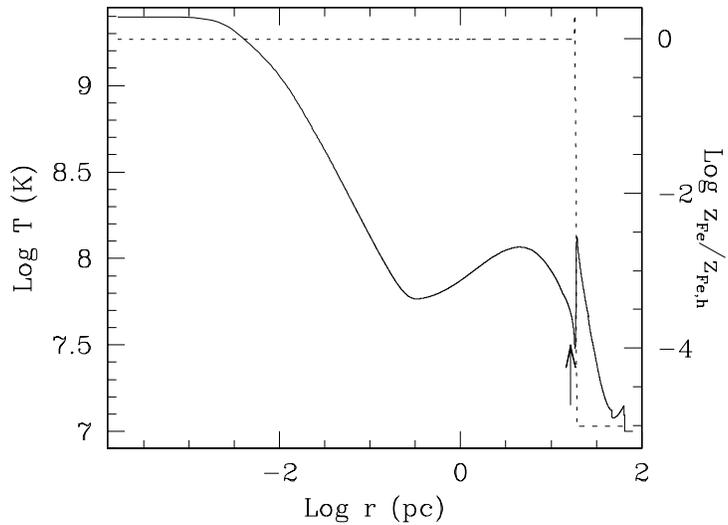}
\vskip.7in
\caption{
Temperature profile ({\it solid line}) of the stellar and
ambient gas produced at time $t = 5 \times 10^4$ yrs 
by a SNIa supernova in hot gas of density $n_e = 0.01$ 
cm$^{-3}$ and temperature $T = 10^7$ K. 
The ({\it dashed}) line shows the computed iron abundance profile 
and the arrow points to the radius that contains $0.7$ $M_{\odot}$, 
the iron core of the SNIa remnant.
}
\label{fig1}
\end{figure}

\clearpage
\begin{figure}
\includegraphics[bb=90 216 522 569,angle=270]
{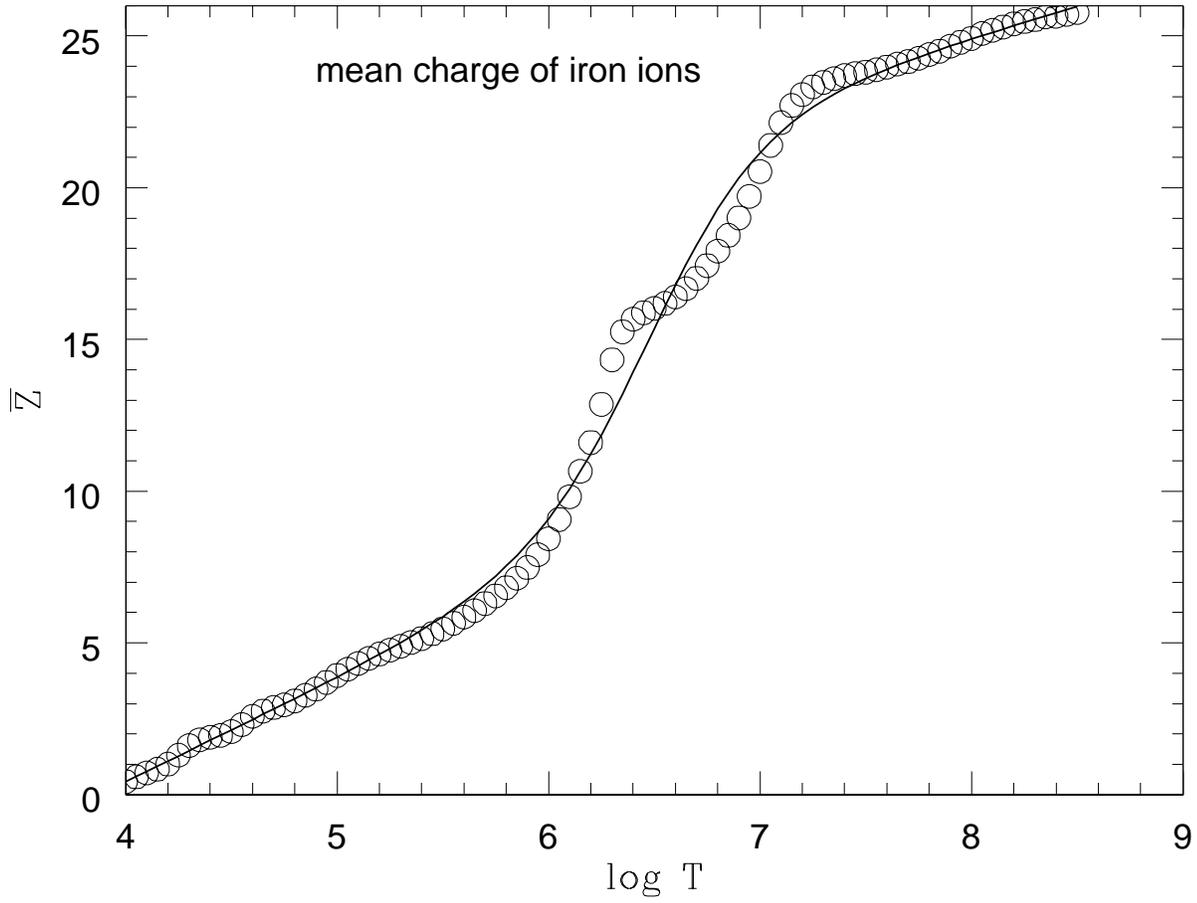}
\vskip.7in
\caption{Mean charge of iron ions ${\bar Z}$ in a collisionally
ionized plasma (Sutherland \& Dopita 1993) together
with an approximate analytic fit.
}
\label{fig2}
\end{figure}

\clearpage
\begin{figure}
\includegraphics[bb=90 216 522 569,angle=270]
{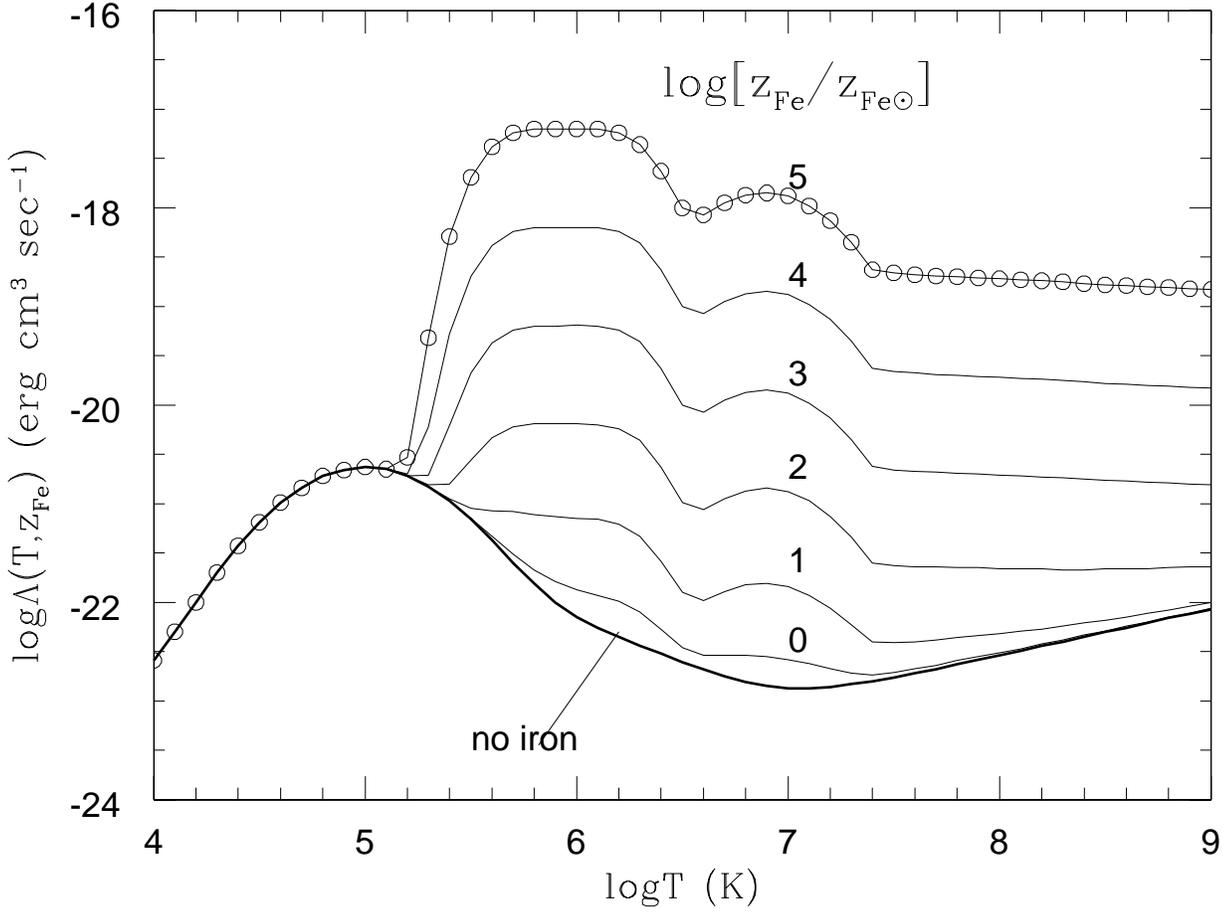}
\vskip.7in
\caption{
Comparison of radiative cooling coefficients without iron 
({\it heavy curve}) and 
with increasing iron abundance. 
Each curve for $\log\Lambda(T,z_{Fe})$ is labeled with
$\log[z_{Fe}/z_{Fe\odot}]$ where
$z_{Fe\odot} = 1.83 \times 10^{-3}$. 
The light solid lines are fits to results 
computed with XSPEC. 
The open circles are the XSPEC results for
$\log[z_{Fe}/z_{Fe\odot}]=5$.
}
\label{fig3}
\end{figure}

\clearpage
\begin{figure}
\includegraphics[width=16cm,angle=0]
{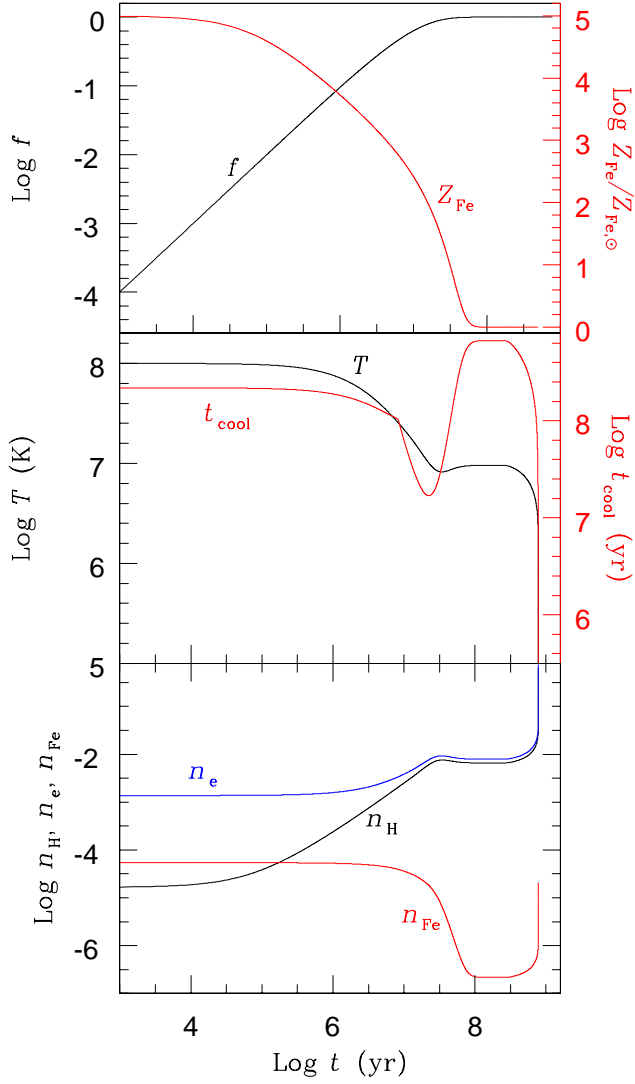}
\vskip.7in
\caption{
Thermal evolution of an iron-rich gas (initially with 
$z_{Fe,h} = 10^5 z_{Fe\odot}$ and $T_h = 10^8$ K) 
mixing with an ambient gas 
(with $z_{Fe,c} = z_{Fe\odot}$ and $T_c = 10^7$ K)
at constant pressure $P_o = 2 \times 10^{-11}$ dyn cm$^{-2}$.
The mixing time is $t_m = 10^7$ years. 
The upper panel shows the fraction $f(t)$ of mixed ambient 
gas and the abundance $z_{Fe}(t)$ of the mixture.
The central panel shows the temperature $T(t)$ and the 
instantaneous cooling time.
The bottom panel shows the number density of electrons, protons 
and iron ions.
}
\label{fig4}
\end{figure}

\clearpage
\begin{figure}
\includegraphics[bb=30 216 462 569,angle=0]
{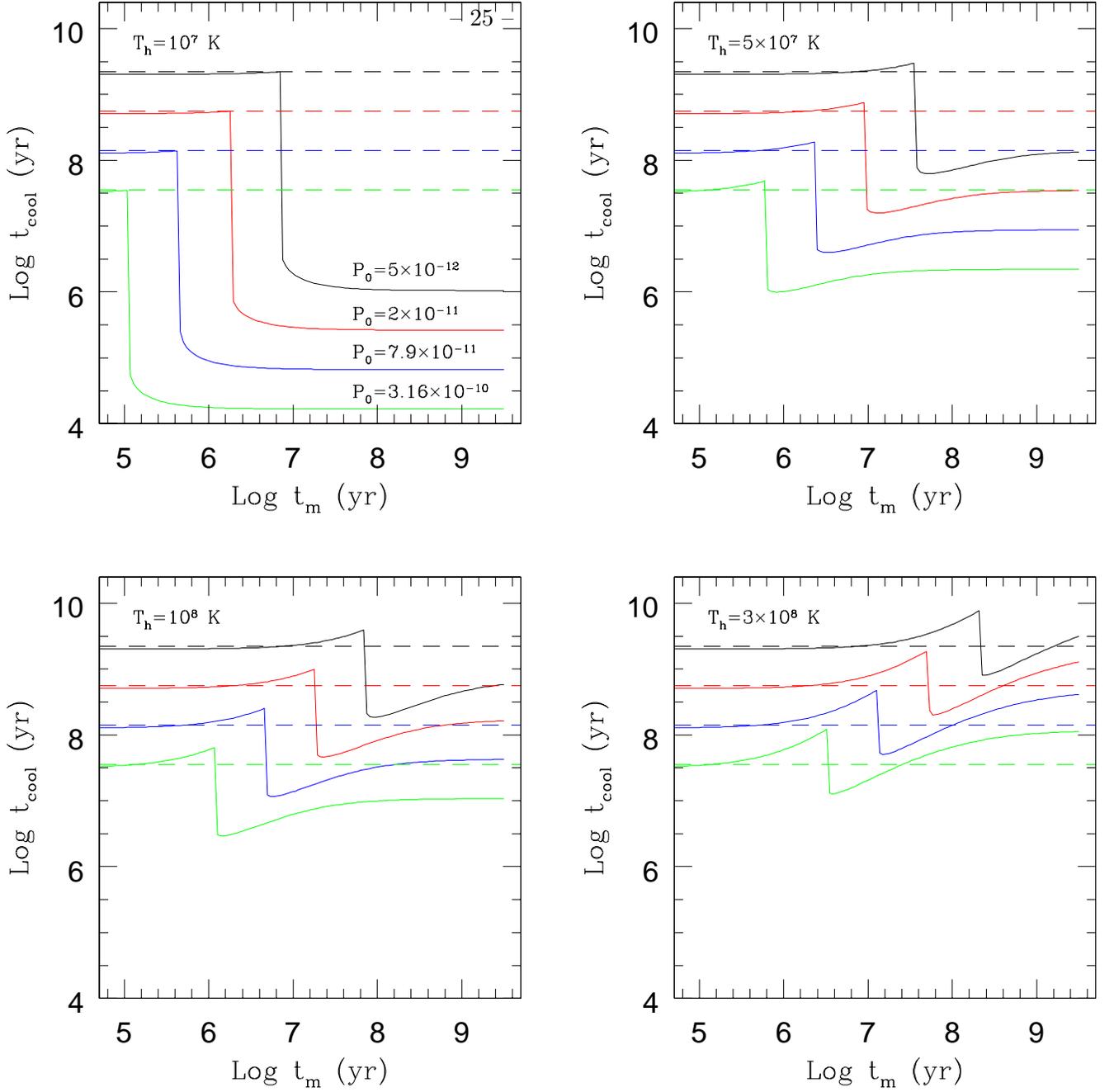}
\vskip.7in
\caption{
Variation of the cooling time (to $T = 10^5$ K) as a function 
of the mixing time $t_m$ with ambient gas at $T_c = 10^7$ K. 
Each panel corresponds to a different initial 
temperature $T_h$ for the hot iron-rich SNIa plasma 
(initially with $z_{Fe,h} = 10^5 z_{Fe\odot}$) and 
the four curves in each panel correspond to four 
pressures $P_o$ (in dynes cm$^{-2}$) 
which remain constant during the cooling.
}
\label{fig5}
\end{figure}

\clearpage
\begin{figure}
\includegraphics[bb=90 216 522 569,angle=0]
{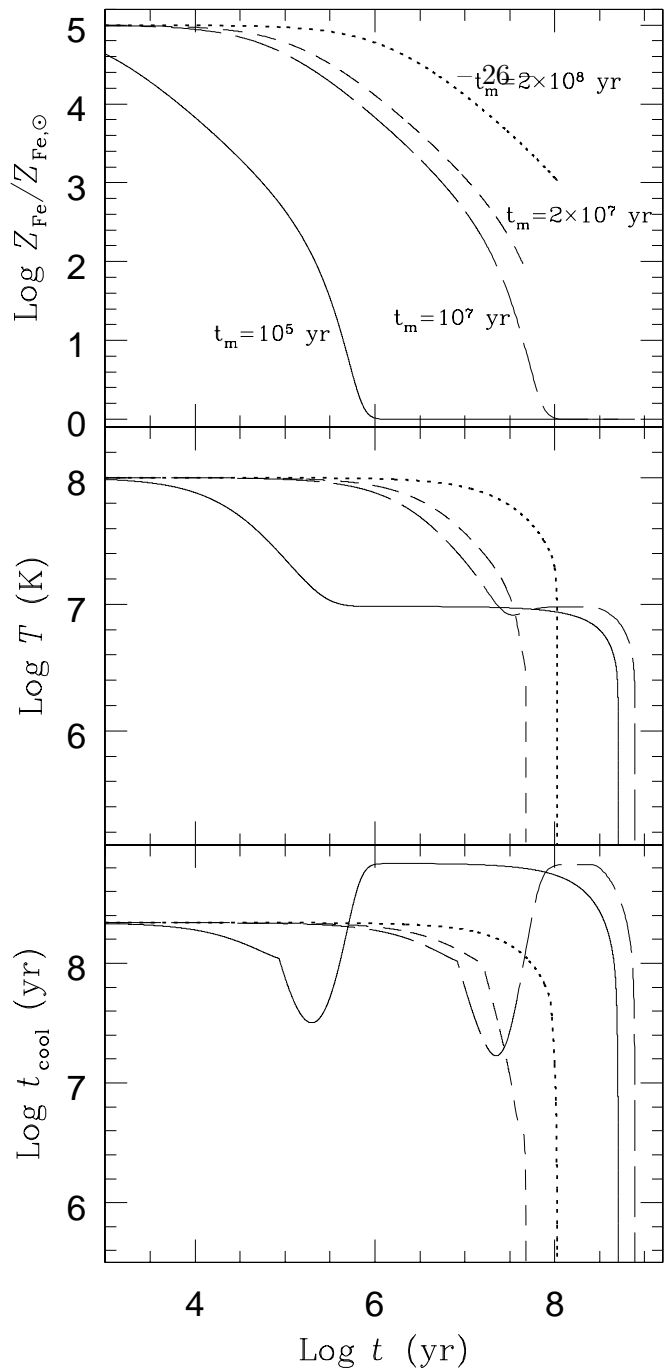}
\vskip.7in
\caption{
Time variation of four iron-rich cooling evolutions corresponding 
to four mixing times 
$t_m = 10^5$ yrs ({\it solid lines}),
$t_m = 10^7$ yrs ({\it long dashed lines}),
$t_m = 2 \times 10^7$ yrs ({\it short dashed lines}) and
$t_m = 2 \times 10^8$ yrs ({\it dotted lines}).
The curves show the 
iron abundance ({\it top panel}),
the temperature ({\it central panel}) and 
the instantaneous cooling time ({\it bottom panel}).
}
\label{fig6}
\end{figure}

\clearpage
\begin{figure}
\includegraphics[bb=90 216 522 569,angle=0]
{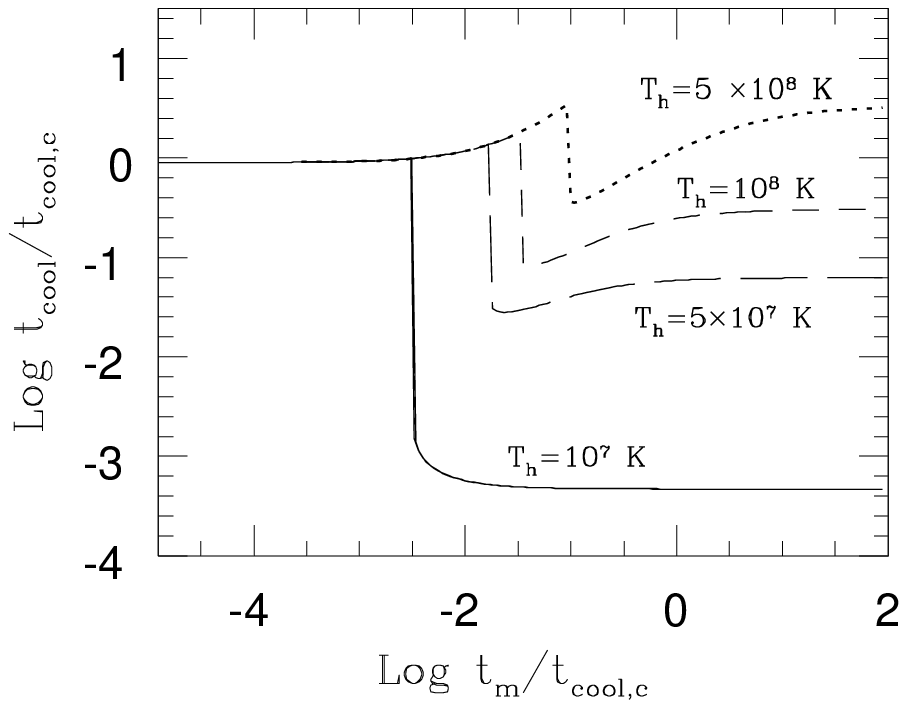}
\vskip.7in
\caption{
Variation of the cooling time (to $T = 10^5$ K) as a function
of the mixing time $t_m$ with ambient gas at $T_c = 10^7$ K 
with both times normalized to the radiative cooling time 
of the ambient gas $t_{cool,c}$.
These curves are identical to those in Figure 5 and are 
valid for all pressures $P_o$.
}
\label{fig7}
\end{figure}


\begin{references}

\reference{} Baumgartner, W.H., Loewenstein, M., Horner, D.J., 
\& Mushotzky, R.F., 2003, ApJ, submitted (astro-ph/0309166)

\reference{} Blanton, E. L., Sarazin, C. L. \& McNamara, B. R., 2003,
ApJ, 585, 227



\reference{} Brighenti, F., Mathews, W. G. 1999, ApJ, 515, 542

\reference{} Buote, D. A., Brighenti, F. \& Mathews, W. G., 
2004, ApJ, 607, L91

\reference{} Buote, D. A., Lewis, A. D., Brighenti, F. \& Mathews,
W. G. 2003, ApJ, 595, 151 {\bf (P10)}



\reference{} Cappellaro, E., Evans, R. \& Turatto, M. 1999, A\&A, 351, 459

\reference{} Chuzhoy, L. \& Loeb, A. 2004, MNRAS, 349, L13


\reference{} Dorfi, E. A. \& V\"olk, H. J. 1996, A\&A, 307, 715

\reference{} Dupke, R. A. \& White, R. E., III, 2000,
ApJ, 537, 123


\reference{} Dwarkadas, V. V. \& Chevalier, R. A. 1998, ApJ, 497, 807


\reference{} Feldman U. 1992, Phys. Scripta, 46, 202

\reference{} Gibson, B. K. \& Matteucci, F. 1997, ApJ, 475, 47



\reference{} Humphrey, P.J., Buote, D.A., \& Canizares, C.R.,
2004, ApJ, in press (astro-ph/0406302)


\reference{} Johnstone, R. M., Allen, S. W., Fabian, A. C. \& Sanders,
J. S., MNRAS, 336, 299


\reference{} Loewenstein, M. 2004, 
Carnegie Observatories Astrophysics Series, vol. 4,
in {\it Origin and Evolution of the Elements},
eds A. McWilliam and M. Rauch (Cambridge: Cambridge Univ. Press),
p. 425



\reference{} Mathews, W. G. 1990, ApJ, 354, 468

\reference{} Mathews, W. G., Chomiuk, L., Brighenti, F. \& Buote,
D. A., ApJ, 616, 745

\reference{} Mathews, W. G., Brighenti, F. \& Buote, D. A. 2004,
ApJ, 615, 662

\reference{} Mathews, W. G., Brighenti, F., Buote, D. A. \&
Lewis, A. D. 2003, ApJ, 595, 159

\reference{} Mathews, W. G. \& Brighenti, F. 2003,
Ann. Rev. Astron. \& Astroph., 41, 191 

\reference{} Narayan, R. \& Medvedev, M. V. 2001, 
ApJ, 562, L129

\reference{} Moretti, A., Portinari, L. \& Chiosi, C., 2003, 
A\&A, 408, 431

\reference{} Morris, R. G. \& Fabian, A. C. 2003,
ApJ, 538, 559

\reference{} O'Sullivan, E. \& Ponman, T. J. 2004, 349, 535


\reference{} Portinari, L., Moretti, A., Chiosi, C. \&
Sommer-Larsen, J., 2004, ApJ, 604, 579


\reference{} Rasmussen, J., Ponman, T.J., 2004, MNRAS, 349, 722

\reference{} Renzini, A., 1997, ApJ, 488, 35

\reference{} Sanders, J. S. \& Fabian, A. C., 2002,
MNRAS, 331, 273



\reference{} Schmidt, R. W., Fabian, A. C. \& Sanders, J. S., 2002, 
MNRAS, 337, 71


\reference{}  Spergel, D. N., et al. 2003, ApJS, 148, 175


\reference{} Sutherland, R. S. \& Dopita, M. A. 1993,
ApJS, 88. 253

\reference{} Tamura, T. et al. 2004, A\&A 420, 135

\reference{} Tornatore, L., Borgani, S., Matteucci, F., 
Recchi, S. \& Tozzi, P., 2004, MNRAS, 349, L19


\reference{} Tozzi, P., Rosati, P., Ettori, S., Borgani, S., 
Mainieri, V., \& Norman, C., 2003, ApJ, 593, 705

\end{references}
\end{document}